\documentclass[12pt]{article}  
\font\titlefont=cmbx10 scaled \magstep4

\def\gprox{\mathrel{\raise .3ex\hbox{$>$\kern-  
.75em\lower1ex\hbox{$\sim$}}}}

\begin{document}   
\input{epsf}   
   
\begin{flushright}   
\vspace*{-2cm}  
gr-qc/0009076 \\   
March 23, 2001 \\    
\vspace*{1cm}   
\end{flushright}   
   
\begin{center}   
{\titlefont CLASSICAL SCALAR FIELDS \\   
\vskip 0.2in   
AND \\   
\vskip 0.2in   
 THE GENERALIZED SECOND LAW} \\   
\vskip .7in   
L.H. Ford\footnote{email: ford@cosmos.phy.tufts.edu}   
 and    
Thomas A. Roman\footnote{Permanent address: Department of Physics and Earth   
Sciences,\\   
Central Connecticut State University, New Britain, CT 06050 \\   
email: roman@ccsu.edu} \\   
\vskip .2in   
Institute of Cosmology\\   
Department of Physics and Astronomy\\   
Tufts University\\   
Medford, Massachusetts 02155\\   
\end{center}

\vspace*{1cm}   
\begin{abstract}   
It has been shown that classical non-minimally    
coupled scalar fields can violate all of the standard    
energy conditions in general relativity. Violations of the    
null and averaged null energy conditions obtainable with such    
fields have been suggested as possible exotic matter candidates    
required for the maintenance of traversable wormholes. In this    
paper, we explore the possibility that if such fields exist, they   
might be used to produce large negative energy fluxes and   
macroscopic violations of the generalized second law (GSL) of   
thermodynamics. We find that it appears to be very easy to produce   
large magnitude negative energy fluxes in flat spacetime.   
However we also find, somewhat surprisingly, that these same types   
of fluxes injected into a black hole do {\it not} produce   
violations of the GSL. This is true even in cases   
where the flux results in a decrease in the area of the horizon.   
We demonstrate that two effects are responsible for the rescue   
of the GSL: the acausal behavior of the horizon and the   
modification of the usual black hole entropy formula by an   
additional term which depends on the scalar field.  
\end{abstract}   
\newpage

\baselineskip=14pt   
\def\eff{{\mathrm{eff}}}    
\section{Introduction}   
\label{sec:intro}   
The weak energy condition (WEC) states that \cite{HE}:    
$T_{\mu\nu}U^{\mu}U^{\nu} \geq 0$, for all    
timelike vectors, where $T_{\mu\nu}$ is the stress-energy    
tensor of matter, and $U^{\mu}$ is an arbitrary    
timelike vector. By continuity, the    
condition holds for all null vectors as well.    
Physically the WEC implies that the energy density    
seen by all observers is non-negative. All experimentally    
observed forms of classical matter satisfy this condition.    
However, quantum matter fields can violate this and all the    
other standard energy conditions of general relativity \cite{EGJ}.    
In at least some circumstances, quantum field theory imposes    
certain restrictions on the degree of energy condition    
breakdown, in the form of what have come to be known as    
``quantum inequalities''. These typically restrict the    
magnitude and duration of negative energy densities and fluxes,    
and seem to limit the production of gross macroscopic effects.    
The quantum inequalities put severe restrictions    
on the realizability of traversable wormholes    
and warp drives \cite{MT,MTY,A,FRWH,PFWD}.    
 
It is often assumed that all classical fields will obey the WEC. This is, 
however, not always the case. The energy density for a classical 
non-minimally coupled scalar field can in fact be negative.  
To our knowledge, the associated stress tensor was first written 
down for the case of conformal coupling by Chernikov and Tagirov~\cite{CT68}. 
The advantages of conformal coupling were emphasized by Callen, Coleman, 
and Jackiw~\cite{CCJ70} for flat spacetime quantum field theory, and further 
discussed by Parker~\cite{Parker} in curved spacetime.  
Bekenstein~\cite{Bekenstein-transf,Bekenstein_PR} noted that the conformal 
scalar field can violate the WEC and other energy conditions, and took 
advantage of this property to construct non-singular cosmological 
models. The violation of energy conditions by non-minimal scalars was  
further discussed by many authors, including Deser \cite{Deser},   
Flanagan and Wald \cite{flanagan}, and in  more detail recently by  
Barcelo and Visser \cite{BV1,BV2,BV}. 
 These features have enabled Barcelo and Visser     
to construct macroscopic wormhole solutions in which non-minimally    
coupled scalar fields violate the averaged null energy condition and 
 serve as the ``exotic matter'' source required    
for wormhole maintenance. Because such fields are classical,    
they are not subject to the quantum inequality restrictions    
on the magnitude and duration of negative energy.    
   
This situation is rather unsettling, since the door    
is then opened for all sorts of bizarre effects. One of    
the most disturbing of these is a potential violation of    
the second law of thermodynamics by creating fluxes    
of negative energy. For example, one might shine such a flux    
at a hot object and decrease its entropy. If the radiation    
field has zero entropy, then the second law will be violated.    
In the case of negative energy fluxes produced by quantum fields,    
the quantum inequalities appear to prevent such large scale    
breakdowns of the second law \cite{F78,F91}. Indeed in some    
cases, such as the Hawking evaporation of black holes,    
negative energy is required for the consistency of the    
unification of the laws of black hole physics and the    
laws of thermodynamics. The generalized second law (GSL) states    
\cite{Bekenstein-GSL} that the sum of the entropy of a black hole   
and that of any  surrounding matter can never decrease, so    
\begin{equation}   
\Delta S_{total} = \Delta S + \Delta S_{matter} \geq  0 \,    
\label{GSL}   
\end{equation}   
where here $S$ is taken to be the entropy of the black hole,    
which is proportional to the area of its event horizon. In    
the Hawking evaporation process, although the area and thus    
the entropy of the black hole decrease, this is more than    
compensated for by the entropy of the emitted    
thermal Hawking radiation.    
   
In this paper we demonstrate that large, transient negative   
energy fluxes can be produced quite easily with   
classical free massless non-minimally coupled    
scalar fields, even in flat spacetime. Such fluxes appear to    
have magnitudes large enough to violate the second law for arbitrary  
lengths of time.  However,  due to possible uncertainties of how   
such fields might interact    
with ordinary matter, one could conceivably argue that perhaps    
the second law might still hold. In an attempt to circumvent the    
latter possibility, we examine a classical non-minimally coupled    
scalar field on a black hole background. We show    
that although the integrated energy flux injected into the hole   
is positive, it can be made  temporarily negative, and that these   
periods can be made arbitrarily long. This raises the possibility   
that one might be able to achieve temporary but large, and in   
principle measurable, violations of the GSL.   
We give a proof that   
the GSL is in fact always {\it satisfied}, even though the horizon area   
of the black hole can temporarily decrease. A physical understanding   
of how this happens involves the subtle interplay of two effects -   
the acausal nature of the horizon and an additional term in the black hole   
entropy which depends on the scalar field. Two illustrative   
numerical examples are presented which demonstrate that a consideration   
of both effects is crucial to the preservation of the GSL.   
We conclude with a summary of our results and a discussion of open questions.   
    
In this paper, we use the MTW metric signature and sign conventions, and    
work in units where $G=\hbar=c=1$.   
   
\section{Negative Energy Fluxes}   
\label{sec:NEF}   
The Einstein equations for a generically coupled scalar field    
can be obtained by varying the Einstein--Hilbert action:     
  \begin{equation}    
  {\cal S}    
  =    
  \frac{1}{16 \pi} \int \;d^4x\sqrt{-g}\;  R    
  +\int \;d^4x\sqrt{-g}    
  \left(    
  -\frac{1}{2} g^{\mu\nu} \; \partial_{\mu}\phi \; \partial_{\nu}\phi    
  -V(\phi)-{1 \over 2}\xi \;R \;\phi^2    
  \right)\,.    
  \label{eq:nlagrangian}    
  \end{equation}    
 The resulting stress-energy tensor,    
${\cal T}_{\mu\nu}$, for the scalar field has the form    
  \begin{eqnarray}    
  {\cal T}_{\mu\nu}    
  =&&\hspace*{-6mm}    
  \nabla_{\mu} \phi \; \nabla_{\nu} \phi    
  - {1 \over 2} g_{\mu\nu} (\nabla \phi)^2    
  - g_{\mu\nu} \; V(\phi)    
  \nonumber \\    
  &&\hspace*{-6mm}    
  +\xi\left[ G_{\mu\nu} \; \phi^2    
  - 2 \;\nabla_{\mu} ( \phi \; \nabla_{\nu} \phi)    
  + 2 \; g_{\mu\nu} \; \nabla^{\lambda} (\phi \; \nabla_{\lambda} \phi)    
  \right],    
  \label{eq:emt}    
  \end{eqnarray}    
where $G_{\mu\nu}$ is the Einstein tensor, and $V(\phi)$ is the scalar    
potential. If all the dependence on $G_{\mu\nu}$ is grouped on the    
left hand side of the Einstein equations, then we can rewrite them 
 by using an effective energy-momentum   
tensor given by    
  \begin{eqnarray}    
  T_{\mu\nu}    
  =&&\hspace*{-6mm}    
  {1 \over 1-  8 \pi \xi \phi^2}    
  \bigg[    
  \nabla_{\mu} \phi \; \nabla_{\nu} \phi    
  - {1 \over 2} g_{\mu\nu} (\nabla \phi)^2    
  - g_{\mu\nu} \; V(\phi)    
  \nonumber \\    
  &&\hspace*{-6mm}    
  -2\, \xi\left[    
   \;\nabla_{\mu} ( \phi \; \nabla_{\nu} \phi)    
  -  \; g_{\mu\nu} \; \nabla^{\lambda} (\phi \; \nabla_{\lambda} \phi)    
  \right]    
  \bigg].    
  \label{eq:emt2}    
  \end{eqnarray}    
This is the expression used by Barcelo and Visser for their analysis of energy    
condition violations. The Einstein equations now read    
$G_{\mu\nu} = 8 \pi T_{\mu\nu}$. Thus a constraint on this    
effective stress-energy tensor is translated directly into a    
constraint on the spacetime curvature. Note that $T_{\mu\nu}$ will become 
singular as $(1-  8 \pi \xi \phi^2) \rightarrow 0$, unless the numerator  
in Eq.~(\ref{eq:emt2}) vanishes at least as rapidly.  
In this paper, we will make the restriction that  
$(1- 8 \pi  \xi \phi^2) > 0$ everywhere. The wormhole solutions of    
Barcelo and Visser have the feature that $(1- 8 \pi  \xi \phi^2)$ can be 
negative, which implies either  super-Planckian values for    
$\phi$ or extremely large values of the coupling parameter $\xi$.    
However, we will show that even with the restriction that    
$(1- 8 \pi  \xi \phi^2) > 0$, one can produce negative energy fluxes   
of large magnitude.

\subsection{Fluxes in Flat Spacetime}   
\label{sec:FFST}   
In the remainder of this section we will set $V(\phi)=0$. Let us first    
demonstrate that the effective stress-tensor, Eq.~(\ref{eq:emt2}),    
can lead to potential problems with the second law even in flat spacetime.    
We choose the simple case of waves travelling only in the positive   
$x$-direction, where    
\begin{equation}   
\phi=\phi(t-x) \,.   
\label{eq:phiflat}   
\end{equation}   
 In this case, the energy    
density $T^{tt}$, and the flux $T^{tx}=F_x$, are equal, and    
so the energy density will also be negative when the flux is negative.    
(To distinguish the case of a positive flux in the minus    
$x$-direction from a true negative energy flux, we count the    
flux as negative only when the energy density is simultaneously    
negative.)  We will assume in this subsection that    
\begin{equation}   
 8 \pi \xi \phi^2  \ll 1 \,.   
\label{eq:kappagg}   
\end{equation}   
   
With the above choices, the energy flux in flat spacetime reduces to    
\begin{equation}   
F_x = T^{tx}= -T_{tx}=   
[(1 - 2\xi)(\phi')^2 - 2\xi \phi \phi''] = (\phi')^2 -\xi (\phi^2)'' \,,   
\label{eq:fluxFST}   
\end{equation}   
where we have used the fact that if $\phi$ has the form of    
Eq.~(\ref{eq:phiflat}),    
\begin{eqnarray}   
&\partial_t \phi & = -\partial_x \phi = \phi' \nonumber \\   
&\partial_x \partial_t \phi &= - \phi'' \,.   
\end{eqnarray}   
Let $t_0$ be a point at which $\phi'$ has a nonzero extremum, so that    
$\phi'(t_0)=0$, but $\phi \neq 0$ and $\phi'' \neq 0$. Then, we can    
make $F_x < 0$ in a neighborhood of $t_0$ by arranging to have  
$- 2\xi \phi \phi'' < 0$ at $t_0$. 
   
As an example, choose $\xi>0$ and    
\begin{equation}   
\phi(t-x)= B \, {\rm sin} \, \omega(t-x) + b \,,   
\label{eq:Asin}   
\end{equation}   
where $\omega=k=k_x>0$, and we assume that $B>0,\,b>0$. Require that    
\begin{equation}   
{1 \over { 8 \pi \xi}} \gg (B+b)^2 \,,   
\label{eq:kappacond1}   
\end{equation}   
which will satisfy our condition Eq.~(\ref{eq:kappagg}). 
 The instantaneous flux, when    
${\rm cos} \, \omega (t-x) =0$ and ${\rm sin} \, \omega (t-x) = -1$, is    
\begin{equation}   
F_x = -2 \xi B^2 \omega^2 \biggl( {b \over B} -1 \biggr) \,.   
\label{eq:flatFinst1}   
\end{equation}    
Let $B^2 = \beta / \xi$, with $\beta \ll 1$    
and $b/B-1 \sim O(1)$, which will satisfy    
Eq.~(\ref{eq:kappacond1}). Then    
\begin{equation}   
F_x \approx  -2 \beta \, \omega^2 = - \beta \,\frac{8 \pi^2}{T^2} \, ,  
\label{eq:flatFinst2}  
\end{equation}  
where $T= 2\pi/\omega$ is the period of oscillation. This expression is in   
Planck units. We can convert to conventional units by recalling that the  
Planck unit of flux is  
\begin{equation}  
F_{p} = \frac{m_p}{l_p^2\, t_p} \approx  
 1.6 \times 10^{104} {\rm {g \over {cm^2 \, s}}} \, ,  
\end{equation}  
where $m_p \approx 2.2 \times 10^{-5} {\rm g}$, $l_p \approx 1.6  
 \times 10^{-33} {\rm cm}$, and $t_p \approx 5.4 \times 10^{-44} {\rm s}$  
are the Planck mass, Planck length and Planck time, respectively. Then  
Eq.~(\ref{eq:flatFinst2}) becomes  
\begin{equation}  
F_x \approx - 8 \pi^2 \,\beta \, F_p\, \frac{t_p^2}{T^2}   
\approx -3.7 \times    
10^{19} \, {\rm {g \over {cm^2 \, s}}} \,\,    
\beta \,\,\biggl({1 {\rm s} \over T } \biggr)^2\,.   
\label{eq:flatFinst2b}   
\end{equation}    
If, for example, we set $T=1$s, $b=2B$ and $\beta = 0.01$,  we have a   
negative energy flux of magnitude    
\begin{equation}   
|F_x| \approx 3.7 \times 10^{17} \,    
{\rm {g \over {cm^2 \, s}}} \,,   
\label{eq:flatFinst3}   
\end{equation}    
which has a duration of the order of a few tenths of a second. By  
ordinary standards, this is an enormous amount of negative energy,  
which has the potential to cause dramatic effects.  
  
Note that Eq.~(\ref{eq:fluxFST}) shows that the flux can be expressed as   
a positive quantity plus a total derivative. This means that if $\phi$  
and its first derivative vanish in both the past and the future, then  
the time-integrated flux is positive. As the above example illustrates,   
however, the transient flux can be made negative for an arbitrarily long   
time.  
  
In order to argue that this flux can be used to achieve a violation    
of the second law, we need to specify how the scalar field interacts    
with ordinary matter. It might be argued that if this interaction    
is sufficiently weak, then second law violations might be avoided.    
Without a detailed model of a system to detect the negative flux,    
this question may be hard to answer definitively. Our purpose here    
was to demonstrate that one can already achieve negative energy fluxes    
of large magnitude, which have the {\it possibility} of violating    
the second law, even in flat spacetime. In the next section we    
turn our attention to a relatively unambiguous energy flux    
detector - a black hole.   
   
\subsection{Fluxes in Curved Spacetime}   
\label{sec:FCST}   
In this section, we will consider the absorption of    
a classical scalar negative energy flux by a black    
hole. In Schwarzschild spacetime, from Eq.~(\ref{eq:emt2}),    
a flux in the radial direction is given by    
\begin{equation}   
T^{tr}=-T_{tr}= -{1 \over 1- 8 \pi \xi \phi^2} \,   
\biggl[ (1-2 \xi) \phi,_t \phi,_r- 2 \xi    
\biggl(\phi \phi,_{tr} - {C,_r \over 2 C} \, \phi \phi,_t   
\biggr)\biggr] \,,   
\label{eq:Ttr}   
\end{equation}   
where $C=1-2M/r$, and $C,_r=2M/r^2$.  If we redefine the    
radial coordinate in the standard way, using $r^*=\int C^{-1} dr$,    
we can express the right hand side as    
\begin{equation}   
{T_t}^r = {{1 } \over 1- 8 \pi \xi \phi^2} \,   
\biggl[ (1-2 \xi) \phi,_t \phi,_{r^*} - 2 \xi    
\biggl(\phi \phi,_{t {r^*}} - \frac{1}{2}  C,_r \, \phi \phi,_t   
\biggr)\biggr] \,,   
\label{eq:Ttr2}   
\end{equation}   
Consider an ingoing wave of the form 
\begin{equation} 
\phi=\phi(v,\theta,\varphi)\,, 
\end{equation} 
where $v=t+r^*$ and $(\theta,\varphi)$ are angular coordinates. Then 
\begin{equation} 
\phi,_t = \phi,_{r^*} = \phi,_v \,, 
\end{equation} 
and 
\begin{equation}   
{T_t}^r = {{1 } \over 1- 8 \pi \xi \phi^2} \,   
\biggl[ (1-2 \xi) (\phi,_v)^2  - 2 \xi    
\biggl(\phi \phi,_{vv} - \frac{1}{2}  C,_r \, \phi \phi,_v   
\biggr)\biggr] \,,   
\label{eq:Ttr3}   
\end{equation} 
 
The mass change of the black hole due to the absorption    
of the flux can be calculated from   
\begin{equation}   
E= \int T_{\mu\nu}\, {\beta}^{\mu} \, d{\Sigma}^{\nu} \,,   
\label{eq:E}   
\end{equation}   
where ${\beta}^{\mu}$ is the timelike Killing vector and    
$d{\Sigma}^{\nu}$ is the area element of the three-dimensional    
hypersurface defined by the black hole's horizon. Here we are    
assuming that the fractional change in the black hole's mass,    
$M$, is small over the time scale $M$, the light travel time    
across the hole. Therefore the metric is approximately    
Schwarzschild and hence has a timelike Killing vector. By    
energy conservation, the time rate of change of the mass    
is given by \cite{FR92}   
\begin{equation}   
\dot{M} = {dM \over dt} = F = \int {T_t}^r \, r^2 d\Omega\,.   
\label{eq:Mp}   
\end{equation}  
On the horizon, ${T_t}^r = T_{vv}$, so this becomes    
\begin{eqnarray}   
\dot{M} &=& 4 M^2 \int \,d\Omega \,{({T_t}^r)}_{r=2M} \nonumber \\   
&=& 4 M^2 \, \int \,d\Omega \,   
{\biggl\{ {1  \over (1-  8 \pi \xi \phi^2)} \,   
\biggl[ (1-2 \xi) {(\phi,_v)}^2 - 2 \xi    
\phi \phi,_{vv} + {\xi \over 2M} \, \phi \phi,_v \biggr]    
\biggr\} }_{r=2M}  \nonumber \\  
&=& \frac{M^2}{2 \pi} \, \int \,d\Omega \, 
\left\{ \left[\ln\left(1 -8 \pi \xi \phi^2\right)\right]_{,vv}  
- \frac{1}{4M}\, \left[\ln\left(1 -  
                               8 \pi \xi \phi^2\right)\right]_{,v}  
 \right\} \nonumber \\  
&+& {\rm a\; positive\; quantity} \,,   
\label{eq:MpH}   
\end{eqnarray}   
where we have used the fact that, on the horizon, $C,_r=1/(2M)$.   
As an example, if we initially assume that $\xi >0$ and $\phi > 0$,    
then when $\phi,_v=0$ and $\phi,_{vv} > 0$, we can have $\dot{M} < 0$.    
   
Strictly speaking,    
$r^* \rightarrow -\infty$ at $r=2M$.    
However, we can get around this problem by taking the effective    
boundary of the hole to be at $r=2M+\epsilon$. That is, we draw a    
surface very close to the horizon, at a finite value of $r^*$,    
and assume that any waves that pass through this surface also go    
into the black hole. Then we have 
\begin{equation}   
\dot{M} = {dM \over dt} = {dM \over dv} \,. 
\end{equation}   
As is the case for the flat spacetime flux, $\dot{M}$ can be written as  
the sum of a total derivative with respect to $v$ and a positive quantity.  
Thus if $\phi$ and $\phi,_v$ vanish as $ v \rightarrow \pm \infty$, the  
net change in $M$ is positive. However, one might worry that this still  
allows the possibility of macroscopic violations of the second law over  
long periods of time.

\section{Black Hole Entropy}   
\label{sec:Entropy}   
It has been shown by a number of authors~\cite{W,VDBH1,VDBH2} that    
in the presence of a classical field on the horizon, the    
entropy of a black hole is not simply $S=A/4$, where $A$ is the   
area of the event horizon, but is modified by   
the presence of additional terms.   
Visser \cite{VDBH1} has referred to these and similar objects as ``dirty''    
black holes. For a classical non-minimally coupled scalar field,    
the Lagrangian density is given by  
\begin{equation}   
L = -{1 \over 2} \bigl( g^{\mu\nu} \, \nabla_{\mu} \, \phi \,   
\nabla_{\nu} \, \phi + \xi \, R \, {\phi}^2 + V(\phi)\bigr)\,.   
\label{eq:Lphi}   
\end{equation}   
In the case where the Lagrangian is an arbitrary function of the    
scalar Ricci curvature $R$, the entropy of a stationary  
black hole is given by (see Eq. (59) of Ref. \cite{VDBH1})    
\begin{equation}   
S = {A \over 4} + 4 \pi \, \int_H \,    
{ {\partial {\cal L}} \over {\partial R} } \,  \sqrt{{}_2 g} \,d^2 x \,,   
\label{eq:Sgen}   
\end{equation}   
where $\cal L$ is the (Euclidean) Lagrangian density for the    
scalar field, and where we have set $\hbar = k =   
$Boltzmann's constant$= 1$. The integration runs over a spacelike    
cross section of the horizon $H$. In this section, our arguments are  
not restricted to Schwarzschild black holes.   
   
In our case, ${\partial \cal L} / {\partial R}   
= -(1/2) \xi \, {\phi}^2$, so  
\begin{equation}   
 \int_H \,    
{ {\partial {\cal L}} \over {\partial R} } \, \sqrt{{}_2 g} \,d^2 x  \,   
= \, -{1 \over 2} \, \xi \, \langle{\phi}^2 \rangle \, A \,,   
\end{equation} 
where  
\begin{equation} 
\langle{\phi}^2 \rangle = \frac{1}{A} \, \int \phi^2 \, \sqrt{{}_2 g} \,d^2 x  
\label{eq:avgphi} 
\end{equation}  
is the average of $\phi$ over the 
horizon. The resulting expression for the entropy of a stationary  
black hole is then simply    
\begin{equation}   
S= {A \over 4} \, ( 1- 8 \pi \xi  \, \langle {\phi}^2 \rangle ) \,.   
\label{eq:S}   
\end{equation}   
One can interpret Eq.~(\ref{eq:S}) as    
\begin{equation}   
S= \frac{A}{4 \,G_{eff}} \,,   
\label{eq:Seff}   
\end{equation}   
where    
\begin{equation}   
G_{eff}=( 1- 8 \pi \xi  \, \langle {\phi}^2 \rangle )^{-1} \,.    
\label{eq:Geff}   
\end{equation}   
The scalar field effectively modifies the local value of Newton's    
constant. When $\xi=0$, the case of minimal coupling, our    
expression reduces to the usual Bekenstein-Hawking entropy. The  
second term only contributes to the black hole's    
entropy when $\phi \neq 0$ on the horizon. Here we are    
discussing the effect of the presence of the scalar field    
on the entropy of the black hole. Since the radiation field itself    
is a particular solution of a classical wave equation, 
 its entropy $S_{matter}$ is zero.   
   
Initially we assume that $\phi =0$ on the horizon, so the    
initial entropy of the black hole is $S_0 = A/4$. We then    
shine in a classical scalar field flux from infinity onto the horizon.    
During the time that $\phi \neq  0$ on $H$, we get a contribution    
to the black hole entropy from the second term in    
Eq.~(\ref{eq:S}). In order to count as a violation of the GSL,  
any decrease in entropy should be sustainable for a least a time    
of order $M$, in order to make a measurement of the horizon area.    
At a later time, the entropy might increase again due to oscillations    
in the sign of the flux, but we would still have achieved a    
measurable violation of the second law.

\subsection{Einstein vs Jordan Frames}  
 
Here we address the issue of the existence of various  
conformal frames for scalar field  
theories. The form of the action of Eq.~(\ref{eq:nlagrangian})  
 is often called the Jordan frame.   
It is possible to perform a conformal metric   
transformation and field redefinition which   
convert the action into a form in which the  
$\xi\, R\, \phi^2$ term is no longer present. To our knowledge,   
this was first done for the transformation relating a massless   
conformally coupled and a minimally coupled scalar field by   
Bekenstein \cite{Bekenstein-transf}, and later generalized to the case   
of arbitrary values of the coupling parameter by others. (See the excellent    
review article by Faraoni, Gunzig, and Nardone \cite{FGN}  
for details and further references.) Deser~\cite{Deser} pointed out that 
this transformation can remove violations of the WEC. 
The form of the theory in which $\xi$ is transformed to zero is   
called the Einstein frame. The use of the word frame to   
describe these forms of the action is perhaps misleading; the   
transformation between the two is {\it not a coordinate   
transformation}, but rather a field redefinition which mixes   
the nature of the scalar and gravitational fields.   
  
This field redefinition technique has been   
extensively used by Jacobson, Kang, and Myers \cite{JKM}   
to study black hole entropy, particularly in the context   
of higher curvature gravity theories. One could apply their   
technique to the massless non-minimally coupled scalar field case   
as well to argue that if the GSL is preserved in the Einstein frame,   
then it should also be preserved in the Jordan frame. The argument   
goes as follows \cite{J-private}. Each solution in the Jordan frame   
can be mapped, via the field redefinition technique, into a solution   
in the Einstein frame. In particular, the black hole entropy for a   
stationary state, as formulated in the Jordan frame, can be transformed   
to the entropy in the Einstein frame. The latter is just the usual   
Bekenstein-Hawking entropy, $S=A/4$, which we know is non-decreasing   
in the Einstein frame, provided that any matter absorbed by the hole   
satisfies the WEC. Additionally, a non-minimally coupled   
scalar field (which violates the WEC) in the Jordan frame is   
transformed to a minimally coupled scalar field (which obeys the WEC)   
in the Einstein frame. Now suppose that in the Jordan frame the evolution of   
the black hole proceeds from one stationary state to another via a series   
of equilibrium states. Since each solution representing one of those states   
can be transformed into a solution in the Einstein frame, and since in the   
Einstein frame the entropy is a non-decreasing function, it should also be a   
non-decreasing function in the Jordan frame. Hence, the GSL should be   
satisfied in both frames.  
  
The question which now arises is whether the Jordan frame and the  
Einstein frame are physically equivalent to one another. This issue  
has been widely debated in the literature and is reviewed in   
Ref.~\cite{FGN}. There is certainly a formal equivalence; a solution of the 
Einstein-Klein Gordon equations in one frame can be mapped to a solution in  
the other frame. Thus in the Jordan frame, one must have a function which is 
non-decreasing in time obtained by mapping the Einstein frame area to the Jordan 
frame. What is less clear is the physical interpretation that should be attached 
to this function, or what observers in the Jordan frame would actually see if  
negative energy is absorbed by a black hole. Furthermore, one is often interested 
in situations where the full theory is not given by Eq.~(\ref{eq:nlagrangian}) 
alone, but there are additional parts in the action describing other fields. 
The transformation to the Einstein frame would then require transforming 
these parts as well, potentially leading to a very cumbersome formulation 
of the theory. 
 
For these reasons, it is desirable to have a proof of the GSL formulated 
entirely in the Jordan frame. Such proofs have been given by Jacobson, 
Kang, and Myers \cite{JKM}  for gravity theories with actions that are  
polynomials in the Ricci scalar, and by Kang~\cite{KANG} for the Brans-Dicke 
theory. Their methods would presumably also be applicable to the case of  
non-minimal scalar fields. In the next subsection, we will give a somewhat  
different proof of the GSL for this case.

\subsection{Proof of the Generalized Second Law}  
\label{sec:PROOF}

In this section, we will show that the entropy defined by Eq.~(\ref{eq:S}) 
is a non-decreasing function, even when negative energy is being 
injected into the black hole. Strictly speaking, this is the entropy  
of a stationary black hole. Let us now assume   
that the expression given by Eq.~(\ref{eq:S}) is the black hole  
entropy even at each moment during the  
dynamical evolution of the horizon and examine its behavior. 
We assume that $\phi \rightarrow 0$ outside of a finite interval in $\lambda$,
where $\lambda$ is an affine parameter on the horizon.
In general, $\phi$ can vary over the horizon, so 
it is convenient to divide the horizon into $n$ patches, where each patch is  
sufficiently small that there is no spatial variation in $\phi$  
through the interval in $\lambda$ during which $\phi \not= 0$. Thus $S$ 
can be written as a discrete sum 
\begin{equation}  
S= \sum_{i=1}^n \Delta S_i \, , 
\end{equation} 
where 
\begin{equation} 
\Delta S_i = {1 \over 4} \,\Delta A_i \, ( 1- 8 \pi \xi  \, {\phi_i}^2) \,. 
\end{equation} 
Here $\Delta A_i$ is the area of the $i$-th patch and $\phi_i$ is the  
value of $\phi$ on that patch. Now we consider a particular patch, and 
drop the $i$ subscript. The logarithmic derivative of $\Delta A$ with 
respect to $\lambda$ is the expansion, $\theta$  
(see, for example Ref. \cite{WBTH}, p. 137): 
\begin{equation}  
 \frac{1}{\Delta A} \,\frac{d\Delta A }{d\lambda} =  
\frac{d \ln (\Delta A)}{d\lambda} = \theta(\lambda) \,.   
\end{equation}  
Thus we can write 
\begin{equation}  
\frac{d}{d\lambda}\ln(\Delta S) = \theta -  
\frac{8 \pi \xi}{1- 8 \pi \xi \, {\phi}^2}\, \frac{d}{d\lambda} (\phi^2)\,. 
\end{equation} 
We will take another derivative of this equation with respect to $\lambda$ 
and use the Raychaudhuri equation for the null generators of the horizon: 
\begin{eqnarray}  
\frac{d \theta}{d \lambda} &=& -\frac{1}{2} \theta^2 -\omega^2 -\sigma^2  
- R_{\mu \nu} k^{\mu} k^{\nu}  \nonumber  \\
&=& -\frac{1}{2} \theta^2 -\omega^2 -\sigma^2  
- 8 \pi \, T_{\mu \nu} k^{\mu} k^{\nu} \,,  
\label{eq:ray1}  
\end{eqnarray}  
where $k^{\mu}$ is the tangent vector to a generator, $\sigma$ is the  
shear and $\omega$ the vorticity.  
  From  Eq.~(\ref{eq:emt2}), we can write 
\begin{equation} 
 T_{\mu\nu} k^{\mu} k^{\nu} = \frac{1}{ 1- 8 \pi \xi \,\phi^2} 
 k^{\mu} k^{\nu} \left[\nabla_{\mu} \phi \; \nabla_{\nu} \phi 
 -2\, \xi \,\nabla_{\mu} ( \phi \; \nabla_{\nu} \phi) \right] \,. 
\label{eq:Tkk} 
\end{equation} 
These relations can be combined to yield 
\begin{equation} 
\frac{d^2}{d\lambda^2}\ln(\Delta S) = -\frac{1}{2} \theta^2 -\omega^2 -\sigma^2 
  - \frac{8 \pi}{1- 8 \pi \xi \,\phi^2}\;  
    \left[1 + \frac{32 \pi  \xi^2 \phi^2}{1- 8 \pi \xi \,\phi^2}\right] \; 
     \left(\frac{d\phi}{d\lambda}\right)^2 \,. \label{eq:d2lnS} 
\end{equation} 
In this calculation, we have used the facts that  
$d/d\lambda = k^{\mu} \nabla_{\mu}$ and that, because $k^{\nu}$ is 
the tangent to a geodesic, $k^{\mu} \nabla_{\mu} k^{\nu}=0$. 
So long as $(1- 8 \pi \xi \,\phi^2) > 0$, all of the terms on the right hand 
side of Eq.~(\ref{eq:d2lnS}) are less than or equal to zero. Hence 
\begin{equation} 
\frac{d^2}{d\lambda^2}\ln(\Delta S) \le 0 \,. \label{eq:lnS_ineq} 
\end{equation} 
 
 Now we wish to show that $\Delta S$ is a monotonically increasing function of 
$\lambda$. First note that $\ln(\Delta S)$ is a monotonic function of 
$\Delta S$, so either one being monotonically increasing implies that 
the other is also monotonically increasing. We assume that the scalar 
field vanishes as $\lambda \rightarrow \infty$, so that $\Delta S \rightarrow 
\frac{1}{4} \Delta A$. Furthermore, at late times $\Delta A$ approaches 
a constant value of $A_f/n$, where $A_f$ is the final horizon area. Thus 
\begin{equation} 
 \frac{d}{d\lambda} \ln(\Delta S) \rightarrow 0 \quad {\rm as} \quad 
    \lambda \rightarrow \infty \,. 
\end{equation} 
If $\Delta S$ were ever to be a decreasing function of $\lambda$, we 
would have to have ${d}[\ln(\Delta S)]/{d\lambda}<\nobreak 0$ at some finite  
value of $\lambda$. However, this is impossible because the only way a 
twice differentiable  
function can go from having a negative value of its derivative to a zero value 
is to pass through a region where the second derivative is positive, which 
is forbidden by Eq.~(\ref{eq:lnS_ineq}). Thus we conclude that 
\begin{equation} 
 \frac{d}{d\lambda}\, \Delta S \geq 0 
\end{equation} 
everywhere.  
Finally, if the entropy contribution of each patch is nondecreasing, it 
follows that the entire entropy is also a nondecreasing function: 
\begin{equation} 
 \frac{d S}{d\lambda}\, \geq 0 \,. 
\end{equation} 
This completes the proof of the generalized second law. Note that $V(\phi)$  
does not appear in Eq.~(\ref{eq:Tkk}), and hence our proof is independent  
of any scalar self-coupling. The proof is not restricted to Schwarzschild  
black holes, and so would also hold for other, e. g., Kerr, black holes  
as well. Here we also did not need to assume that the  
changes in the mass of the black hole must be necessarily small. 
 
We point out that although $S$ is a non-decreasing function, the same is  
not necessarily true of the horizon area.  
If we take the second derivative of $\ln (\Delta A)$, and use the  
Raychaudhuri equation we can write  
\begin{equation} 
\frac{d^2 \ln (\Delta A)}{d{\lambda}^2} 
=-\frac{1}{2} \, {\biggl(\frac{d \ln (\Delta A)}{d\lambda} \biggr)}^2 
-\omega^2 -\sigma^2 - 8 \pi \, T_{\mu \nu} k^{\mu} k^{\nu} \,. 
\label{eq:rayagain} 
\end{equation} 
If $T_{\mu \nu} k^{\mu} k^{\nu} \geq 0$, then  
\begin{equation} 
\frac{d^2 \ln (\Delta A)}{d{\lambda}^2} \leq 0 \,. 
\label{eq:d2Aleq} 
\end{equation} 
However, if the WEC is violated, so that $T_{\mu \nu} k^{\mu} k^{\nu} < 0$,  
then the sign of ${d^2 \ln (\Delta A)}/{d{\lambda}^2}$ can  
be positive, and therefore the argument we made for the non-decrease of $S$  
does not work for $A$. We assumed that the scalar field  
vanishes outside of a finite interval in $\lambda$, so we know that the entropy  
in the distant past is $S_0 = A_0/4$ and in the distant future is  
$S_f=A_f/4$, where $A_0$ and $A_f$ are the initial and final total horizon  
areas, respectively. Since we proved that $S$ is a non-decreasing function,  
this implies that $A_f \geq A_0$. 
The case $A_f = A_0$ corresponds to the trivial case that $\phi = 0$
everywhere, as may be seen from the following argument. 
If $S_f=S_0$, the entropy must be 
constant. However, this can only occur if $\Delta S$ is constant for 
each patch, which in turn can only occur if $\phi = 0$ for all $\lambda$. 
If $\Delta S$ is constant, all of the terms on the right hand side of 
Eq.~(\ref{eq:d2lnS})  must vanish, but the last of these terms can only 
do so for the trivial, $\phi = 0$, case. 
Of course, there is still the possibility of  
temporary decreases in $A$ during the evolution.  
Indeed in the next section, we will exhibit specific examples  
where $A$ can either increase or temporarily decrease, while $S$ is always  
non-decreasing. Note that the fact that $A_f > A_0$ is a generalization of 
our result in the previous section. There we showed that the net change in the
mass of a Schwarzschild black hole is positive in the case of small fractional
changes.

\section{Numerical Examples}  
\label{sec:numex}  
 
In this section, we will construct some numerical examples to illustrate  
the effects of negative energy on a Schwarzschild black hole. We will restrict our  
examples to the case of spherically symmetric pulses which change  
the black hole's mass by a small fraction. 
 
\subsection{General Procedure to Solve for $\theta$ and $A$}  
\label{sec:gen}  
 
The relation between the expansion of the   
null generators, $\theta$, and the area of the horizon, $A$,   
is given by    
\begin{equation}  
\frac{dA}{d\lambda} = A \, \theta(\lambda) \,.  
\label{eq:thetaA}  
\end{equation}  
Strictly speaking, in the general case, $A$ is the area of an   
 infinitesimal cross-sectional area element of a bundle   
of null geodesic generators. The expansion $\theta$ measures the   
 local rate of change of cross-sectional area as one moves   
along the generators. However, in this section, we assume spherically  
symmetric black holes and spherically symmetric   
energy distributions. Therefore the infinitesimal change in horizon   
area will be the same along each bundle of null generators, so   
in our case $A$ can be taken to be the total horizon area.   
  
We also need the relationship of $\lambda$   
to Schwarzschild-like coordinates. For a stationary   
black hole, we can take (see, for example   
Ref. \cite{WBTH}, p. 122)   
\begin{equation}  
\frac{dV}{d\lambda} = \kappa \,,  
\label{eq:Vlambda}  
\end{equation}  
where $\kappa = 1/(4M)$ is the surface gravity   
of the black hole and $M$ is its mass. Here   
\begin{equation}  
V = e^{\kappa v}  
\label{eq:V}  
\end{equation}  
is the Kruskal advanced time.  
Thus we have that   
\begin{equation}  
\lambda = {\kappa}^{-1} V = {\kappa}^{-1} \, e^{\kappa v}\,.  
\label{eq:lambdaV}  
\end{equation}  
We will assume that $T_{vv}(v)$ represents ingoing   
radiation which changes the black hole's mass by only a small   
fractional amount, $|\Delta M| \ll M$. We can then take   
$\kappa$ to be constant to lowest order.  
If we change the independent variable   
from $\lambda$ to $v=t+r^*$, then    
\begin{equation}  
\frac{d}{d \lambda} = e^{-\kappa v} \, \frac{d}{dv} \,,  
\label{eq:dlambdatodv}  
\end{equation}  
and   
\begin{equation}  
T_{\mu \nu} k^{\mu} k^{\nu} = e^{-2 \kappa v} \, T_{vv} \,.  
\label{eq:Tnew}  
\end{equation}

For spherically symmetric pulses, the shear and vorticity vanish, and 
the Raychaudhuri equation, Eq.~(\ref{eq:ray1}), becomes 
\begin{equation}  
\frac{d \theta}{dv} = -\frac{1}{2} \, e^{\kappa v} \,   
{\theta}^2   
- 8 \pi \, e^{-\kappa v} \, T_{vv}(v) \,.  
\label{rayv2}  
\end{equation}  
The equation for the horizon area can be expressed as   
\begin{equation}  
\frac{dA}{dv} = e^{\kappa v} \, A \, \theta \,.  
\label{eq:thetaAv2}  
\end{equation}  
Given an explicit form for $T_{vv}(v)$, we need to   
integrate this pair of equations subject to the final   
conditions that $\theta = 0$ and $A=A_f$ in the future.  
  
The $\delta$-function pulse example presented in the Appendix   
indicates that the horizon tends to anticipate the   
energy flux by a time scale of order ${\kappa}^{-1} = 4M$.   
However, as shown there, the exponential factor $e^{\kappa v}$  
tends to strongly suppress any influence on timescales long compared   
to ${\kappa}^{-1}$. If we send in a negative energy pulse   
with duration much greater than $M$, followed by a   
positive energy pulse, we would expect that $A$ will   
first decrease, and then increase, always a time   
$\sim \, {\kappa}^{-1}$ ahead of the energy pulse.  
  
\subsection{Compact Pulses}  
\label{sec:CP}  
In this section, we construct a specific waveform    
for $\phi$ on the horizon which results in a    
(quasi-periodic) negative energy flux down the black hole.   
In general, for wave propagation in a black hole    
spacetime we can write     
\begin{equation}   
\phi=\phi(t,r^*, \theta, \Phi)= {R(t,r^*) \over r} \, Y_{lm} \,,   
\label{eq:genphi}   
\end{equation}   
where the $Y_{lm}$ are the usual spherical harmonics.    
  
Choose the ingoing modes to be $l=0$ only.   
To have an energy pulse which is bounded in time,   
we will use the following   
form for $\phi$ on the horizon:  
\begin{equation}  
\phi(v) = \frac{v^2 \, {(v-T_0)}^2}{ \,{T_0}^4} \,   
\frac{g_T (t,r^*)}{2 M\,\sqrt{4 \pi}} \,, \quad  0 \le v \le T_0 \,,  
\label{eq:cphi}  
\end{equation}  
and $\phi(v) =0$ if $v < 0$ or $v >  T_0$.  
We will take $g_T$ to be a sum of two   
sinusoidal functions, as this form is an example which   
produces a negative energy  flux down the hole.   
In particular, let    
\begin{eqnarray}   
g_T &=& a_1 \,{\rm cos} \,\omega_1 v +    
a_0 \,{\rm cos} \,\omega_0 v \nonumber \\   
&=& {a_1 \over 2} \, (e^{i \omega_1 v} + e^{-i \omega_1 v}) +    
{a_0 \over 2} \, (e^{i \omega_0 v} + e^{-i \omega_0 v})\,.   
\label{eq:gTspecific}   
\end{eqnarray}   
Note that there would be a potential problem with simply   
choosing $g_T(t,r^*)$ on the horizon to be the form   
given by Eq.~(\ref{eq:gTspecific}) and ``chopping'' it;   
we could introduce $\delta$-function   
terms or their derivatives into $T_{vv}$. To avoid this, we   
chose the form of $\phi$ given by Eq.~(\ref{eq:cphi}), so that   
both $\phi$ and $\phi,_v$ go to zero smoothly at $v=0$ and $v=T_0$.  
  
 We can choose   
$1/M \gg \omega_1 \ge \omega_0$, to get a form    
for $g_T$ that is approximately a constant $+$ a sinusoidal part    
with a frequency much less than $1/M$.  Now let $T_0 = 2\pi / \omega_0$  
and $T_1 = 2\pi / \omega_1$, where   
we choose $T_0 \geq T_1 \gg M$. The latter condition will insure that   
the time periods over which the flux is negative are long compared to $M$.   
This allows the changes in the Schwarzschild geometry to be quasistatic.   
  
The bounds on $a_0$ and $a_1$ are now restricted by the requirement that    
$(1 - 8 \pi \,\xi \, \phi^2) >\nobreak 0$ everywhere. Our pulse is a wavepacket   
which consists of a range of both low and high frequency modes.   
The effect of the potential barrier is to reduce the value of $\phi$   
on the horizon compared to its value just outside the potential barrier.   
The low frequency modes are the ones most   
attenuated by the potential barrier, whereas the high frequency modes   
are the ones least affected. Since $\phi$ increases roughly monotonically   
with decreasing $r$, the maximum value of $\phi$ would occur just outside   
the barrier, at $r = r_0 \approx 4M$. On the other side of the barrier, the   
width of our pulse is $\approx T_0$; hence the wavepacket on the horizon   
will be dominated by modes with frequency $\approx \omega_0$. If we   
trace these modes backwards through the potential barrier, they will grow in   
amplitude by a factor of about ${|{\cal T}_{\omega_0\, 0}|}^{-1}   
\approx {(4 M \, \omega_0)}^{-1}$ where ${\cal T}_{\omega_0\, 0}$ is   
the transmission coefficient for a spherically symmetric wave of 
frequency $\omega_0$. It is given by    
\cite{PFGQI}    
\begin{equation}   
{\cal T}_{\omega_0\, 0}=-4 i M \omega_0 \,,   
\label{eq:T}   
\end{equation}   
for $\omega_0 \ll 1/M$. Note that we need both of the   
frequencies, $\omega_0$ and $\omega_1$, to be nonzero so that the   
transmission coefficients for both components be nonzero. For this reason,  
the choice, Eq.~(\ref{eq:Asin}), used in flat spacetime is not appropriate  
here. We will choose $\xi > 0$, $a_0 > a_1$, and $\omega_1 \gg \omega_0$, so   
\begin{equation}  
|\phi_{max}| \approx \frac{a_0}{64 M \omega_0 \,   
\sqrt{4 \pi}\, r_0} \,,  
\label{eq:phimax}  
\end{equation} 
where we have used the fact that the factor of  
$v^2(v-T_0)^2/T_0^4$ attains its maximum value of $1/16$ at $v=T_0/2$.  
The requirement that $(1 - 8 \pi \xi \, {\phi_{max}}^2) > 0$ gives   
\begin{equation}  
|\phi_{max}| < \frac{1}{\sqrt{8 \pi \xi}} \,,  
\label{eq:phimax2}  
\end{equation}  
so therefore we have that   
\begin{equation}  
a_0 < \frac{256 \pi M^2}{T_0} \,\sqrt{\frac{2}{\xi}} \,,   
\label{eq:a0bound}   
\end{equation}  
where we have used $\omega_0 = 2\pi/T_0$ and $r_0 \approx 4M$.  
  
\subsection{Two Illustrative Examples}  
\label{sec:twoex}  
We now present two interesting numerical examples. The first is a case   
where $A$ never decreases even though the flux is periodically   
negative. In this case, the GSL is rescued   
by the ``precognition'' of the horizon. That is, the horizon   
starts to respond acausally in anticipation of the   
subsequent energy pulse. This counter-intuitive behavior of the   
horizon is discussed in more detail in the Appendix. In the second   
example, the periods of negative energy flux last long enough that   
this precognition effect cannot keep the horizon area from decreasing   
periodically, although the integrated area change is positive. What   
keeps the GSL from being violated during the periods of   
area decrease is the effect of the $\phi^2$ term in the formula for   
the entropy. These examples demonstrate that consideration   
of both effects is important for maintaining the GSL.   
If one or the other of them is left out, one can find examples   
where the law is violated.   
   
Let us choose units in which $M = 1$, so that $\kappa=1/4$.  
Strictly, this is the final mass, but here $|\Delta M|/M \ll 1$, 
so we can assume that $\kappa$ is approximately constant.  
Then our final conditions in $v$ are that   
at $v=T_0$, $\theta=0$ and $A=A_f=16 \pi$.   
  
Case (i): Choose $T_0=40$, $T_1=4$, and $\xi=1/6$. From   
Eq.~(\ref{eq:a0bound}), the bound on $a_0$ in our new units   
is $a_0 < (256 \pi/T_0)\,\sqrt{2/\xi}$, so in this case $a_0 < 69.6$.   
We conservatively choose $a_0=0.16$ and $a_1 = 0.5 a_0$. The   
behavior of the flux, the horizon area, and the entropy are represented   
in Fig. 1. The three quantities have been rescaled in order   
to display all three on the same plot. The plotted rescaled   
quantities are: $F_{scaled} = 3.2 \times 10^6 \, T_{vv}, \,  
A_{scaled} = 10^4 \, (A-A_f)/A_f$, and $S_{scaled}= 10^4 \, (S-S_f)/S_f$, 
where $S_f = A_f/4 =4\pi$ is the final entropy in our units.   
In this case there are three negative energy pulses embedded   
in the overall flux, but the duration of each pulse is short enough   
that the negative and positive parts of the flux occur close   
together in time. The acausal expansion of the horizon,   
in anticipation of the (larger) positive parts of the flux, is   
enough to save the GSL in this case. It is very important to take   
this effect into account in the analysis of the dynamical case.   
For a static black hole, the area of the horizon is simply   
$16 \pi M^2$. This is also the area of the apparent  
horizon in the dynamical case. If one naively adopts this form for the area   
during the dynamical phase (as we mistakenly did in an   
earlier analysis of this problem \cite{W-private}),   
then even taking into account the change in the mass,   
one can still find cases where the GSL appears to be violated.

\begin{figure}   
\begin{center}   
\leavevmode\epsfysize=10cm\epsffile{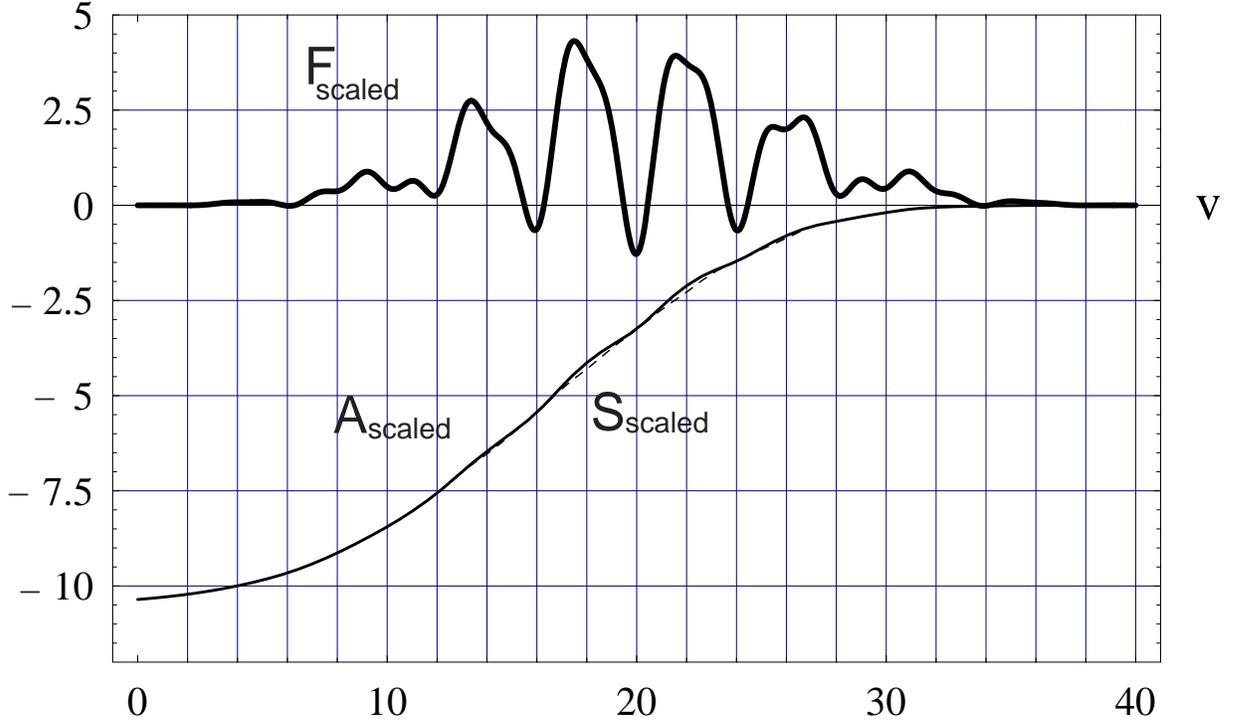}   
\end{center}   
\caption{Flux, horizon area, and black hole entropy as functions of $v$ for   
$T_0=40$, $T_1=4$, $a_0=0.16 = 2 a_1$, and $\xi=1/6$.   
The plotted rescaled   
quantities are: $F_{scaled} = 3.2 \times 10^6 \, T_{vv}, \,  
A_{scaled} = 10^4 \, (A-A_f)/A_f$, and $S_{scaled}= 10^4 \, (S-S_f)/S_f$,   
respectively. The behavior of the horizon area and the entropy are   
almost the same.}   
\label{NoAdecrease}   
\end{figure}   
  
Case (ii): Choose $T_0=400$, $T_1=40$, and $\xi=1/6$. From   
Eq.~(\ref{eq:a0bound}), $a_0 < 6.96$. We again choose   
$a_0=0.16$ and $a_1 = 0.5 a_0$. The behavior of   
the flux, the horizon area, and the entropy are shown   
in Fig. 2. The plotted rescaled   
quantities are: $F_{scaled} = 1.6 \times 10^8 \, T_{vv}, \,  
A_{scaled} = 10^5 \, (A-A_f)/A_f $, and $S_{scaled}= 10^5 \,(S-S_f)/S_f$.   
There are again three negative energy pulses.   
However, in this case, each pulse lasts sufficiently long   
enough that the acausal behavior of the horizon is {\it not}   
sufficient to rescue the GSL, as evidenced by the three   
periodic decreases of the horizon area. That is, if the entropy   
were simply given by $S=A/4$, as in the usual case, then   
the GSL would be violated during the periods of area decrease.   
What prevents this from occurring in this case is the   
influence of the scalar field term in the formula for   
the black hole entropy. Note also the phase lag between the   
change of sign of the flux and the onset of area decrease or   
increase. The scaling factors used in the two figures are different. 
The values of the scalar field on the horizon are approximately the same 
in both cases, and hence the magnitude of the fractional difference 
between $S$ and $A$ are of the same order of magnitude. However, the 
magnitude of the flux is smaller in Fig.~2 than in Fig.~1, because of the  
slower rate of change of $\phi$. 
 
\begin{figure}   
\begin{center}   
\leavevmode\epsfysize=10cm\epsffile{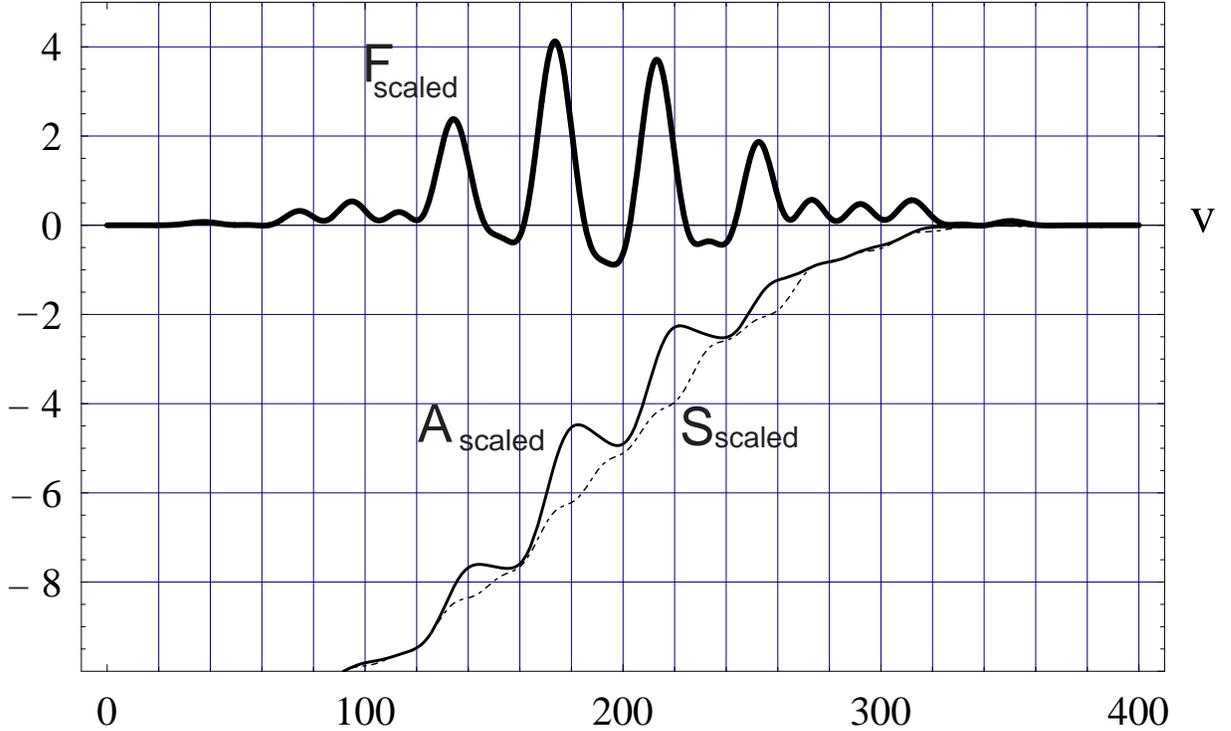}   
\end{center}   
\caption{Flux, horizon area, and black hole entropy as functions of $v$ for   
$T_0=400$, $T_1=40$, $a_0=0.16= 2 a_1$, and $\xi=1/6$.   
The plotted rescaled quantities are: $F_{scaled} = 1.6 \times 10^8 \, T_{vv}, \,  
A_{scaled} = 10^5 \, (A-A_f)/A_f$, and $S_{scaled}= 10^5 \, (S-S_f)/S_f$,   
respectively. Here the horizon area undergoes periodic decreases,   
but the entropy is always non-decreasing.}   
\label{Adecrease}   
\end{figure}

\section{Conclusions}   
\label{sec:conclusions}

We have demonstrated that classical massless non-minimally    
coupled scalar fields can be used to produce disturbingly    
large negative energy fluxes, even in flat spacetime.   
These classical negative energy fluxes are temporary in   
the sense that the time-integrated  
flux is always positive. However, unlike the 
situation in quantum theory, there are no constraints here analogous  
to the quantum inequalities. Therefore, it is possible for   
such fluxes to be sustained over macroscopically long time intervals.  
These fluxes appear to have magnitudes sufficient to    
produce gross violations of the second law of thermodynamics.   
  
However, whether such violations occur could depend on    
the details of how these fields interact with     
matter. In an attempt to circumvent this issue, we   
examined these negative    
energy fluxes in the presence of a relatively unambiguous    
energy detector, a black hole. We considered   
scenarios in which a negative energy flux, composed of classical  
non-minimally coupled scalar fields, is injected into the black hole.   
It was proved that the GSL is in fact always {\it satisfied}. This  
conclusion is true for any positive or negative value of the  
coupling parameter, subject to the condition that $(1-8 \pi \xi \, \phi^2)>0$,  
and is independent of the scalar potential $V(\phi)$. The proof is also not  
restricted to Schwarzschild black holes, and therefore will hold for rotating  
black holes as well. The validity of the GSL hinges on two key effects -   
the acausal behavior of the event horizon and the presence of   
an additional term in the formula for the black hole entropy   
which depends on the scalar field.   
  
In the dynamical case, the area of the event horizon, in contrast 
to that of the apparent horizon, depends 
upon the future history of the spacetime. This behavior is reviewed and 
illustrated in the Appendix. For a Schwarzschild black hole,  
when the temporal separation of the 
negative and positive energy pulses is no more than several times $M$, 
as was the case in Fig.~1, this effect alone can be sufficient to 
preserve the GSL. Here the horizon moves outward in anticipation of 
the (larger) positive energy in such a way that $A$ is non-decreasing. 
 
If, however, the temporal separation becomes long compared to $M$, this  
effect is no longer sufficient to prevent $A$  from decreasing. This 
behavior is illustrated in Fig.~2. A similar result was found in the  
case of Brans-Dicke theory in Ref.~\cite{SST}. Here the $\phi$-dependence 
of $S$ is crucial for insuring that the entropy is non-decreasing.  
Note that this $\phi$-dependence is also not always sufficient by itself 
to save the GSL without the acausal behavior of the horizon. For example, 
the product of the area of the apparent horizon times $(1-8 \pi \xi \, \phi^2)$ 
can decrease.  
 
There are still some questions left unanswered at this point. One is whether 
negative energy could actually destroy a black hole. Our treatment in 
Sect.~\ref{sec:PROOF} and those in Refs.~\cite{JKM,KANG} have assumed 
either a final black hole or cosmic censorship, and hence do not rule out  
this possibility. Another question is whether one could violate the 
GSL if $(1-8 \pi \xi \, \phi^2)$ is not positive everywhere;  
this situation would involve either super-Planckian values of the scalar  
field or very large values of the coupling parameter. It might also result  
in a sign change of the effective local Newton's constant, given by  
Eq.~(\ref{eq:Geff}), and non-positivity of the entropy density. Recall that 
this condition played a crucial role at several steps. We assumed it 
in our proof of the GSL, and it is also needed to define the transformations 
from the Jordan to the Einstein frame, and hence to make the argument that the 
area in the Einstein frame provides a suitable non-decreasing quantity.  
It may also be necessary for the initial value problem to be  
well-posed~\cite{Eanna2}. 
If $(1-8 \pi \xi \, \phi^2) \rightarrow 0$, then one would expect the 
effective stress tensor to become singular. This should lead to a backreaction 
effect which will prevent the system from actually reaching  
$(1-8 \pi \xi \, \phi^2) = 0$; such an effect was indeed found in the case of 
cosmological spacetimes in Ref.~\cite{Saa}. However, $T_{\mu\nu}$ can 
remain finite if the numerator in Eq.~(\ref{eq:emt2}) vanishes at the 
same points that $(1-8 \pi \xi \, \phi^2) \rightarrow 0$. This is what happens 
in some of the solutions of Barcelo and Visser~\cite{BV1,BV2,BV}. Whether one  
could similarly construct examples which violate the GSL  
when $(1-8 \pi \xi \, \phi^2) \leq 0$ remains unknown, although we strongly  
suspect so. Nonetheless, it is quite interesting that, even for  
relatively weak fields and coupling constants: (a) one can get large sustainable,  
albeit temporary, classical negative energy fluxes, and (b) that  
such fluxes do not violate the GSL.  
 
The acausal nature of the area of the event horizon is a disturbing feature  
of the current formulation of black hole thermodynamics, and one which 
has drawn the attention of various authors~\cite{CS}. It is not clear whether  
this feature will survive in  future theories. A definition of entropy 
based on a statistical mechanical enumeration of states would seem to have to  
be causal. This is a deep question which may have to await a more complete 
quantum theory of gravity for a resolution. 
 
Finally, there is the question of whether the large negative energy associated 
with a classical scalar field could produce any dramatic effects, such as 
violations of the second law for an ordinary thermodynamic system. More 
generally, does this negative energy produce any observable effect on 
ordinary matter? It would seem rather peculiar if one could violate the  
ordinary second law but not the GSL. One would suspect that they should  
either both hold or both be violated. However at present we do not  
know of a similar general argument that would guarantee that the ordinary  
second law is not violated.  
 
If such classical scalar fields can exist, that raises the   
question of why only certain classical fields are capable   
of producing large energy condition violations. For example, it is   
often thought that the conformally coupled scalar field is   
``physically reasonable'' because it faithfully mimics certain   
aspects of the electromagnetic field. But why then does the   
classical electromagnetic field obey the WEC, while the conformally   
coupled scalar field does not?  
  
It is usually easier to get negative   
energy in the context of quantum rather than classical fields.  
Moreover, in the regime of quantum fields, there seem to be some   
rather strong restrictions imposed on the magnitude and extent of   
negative energy densities and fluxes, in the form of the quantum   
inequalities. Given that nature seems to tightly restrict quantum 
violations of the weak energy condition, how seriously should one take 
classical violations? These are all issues for further enquiry.

\newpage   
\centerline{\bf Acknowledgments}   
The authors would like to thank Ted Jacobson for stimulating criticisms   
and correspondence, and Bob Wald, Bill Hiscock, and Matt Visser for   
key observations. We also thank Paul Anderson, Carlos Barcelo,  
Arvind Borde, \'Eanna   
Flanagan, Jaume Garriga, and Alex Vilenkin for useful comments and   
discussions. TAR would like to thank the    
members of the Tufts Institute of Cosmology for their    
hospitality while this work was being done. This research    
was supported by NSF Grant No. Phy-9800965 (to LHF)   
and by NSF Grant No. Phy-9988464 (to TAR).

\section{Appendix}  
\label{sec:delta}  
In this appendix we analyze the simple case of a spherical   
$\delta$-function pulse of positive energy imploding onto   
an already-existing black hole. We examine this case for   
two reasons. The first is to illustrate the somewhat   
counter-intuitive dynamical behavior of the horizon which   
is important in our discussion. The second is to   
present this calculation in a pedagogically detailed form -   
one which we have been unable to find in the standard   
literature.  
  
For simplicity, we consider a Schwarzschild black hole  
of initial mass $M_0$, which absorbs a spherically symmetric  
$\delta$-function pulse of positive energy null fluid and becomes  
a black hole of mass $M$. Let $\lambda$ be the affine parameter   
on the horizon and suppose that the pulse intersects   
the horizon at $\lambda = 0$.   
 In the case of a spherically symmetric pulse the Raychaudhuri   
equation for the null generators of the horizon reduces   
to the simple form   
\begin{equation}  
\frac{d \theta}{d \lambda} = -\frac{ {\theta}^2}{2}   
- R_{\mu \nu} k^{\mu} k^{\nu} \,.  
\label{eq:Aray1}  
\end{equation}  
For positive energy $R_{\mu \nu} k^{\mu} k^{\nu}   
= 8 \pi T_{\mu \nu} k^{\mu} k^{\nu} > 0$, so set   
$R_{\mu \nu} k^{\mu} k^{\nu} = a \delta(\lambda)$,   
where $a$ is a positive constant.   
  
Let us first solve the Raychaudhuri equation without   
the $\delta$-function term:  
\begin{equation}  
\frac{d \theta}{d \lambda} = -\frac{ {\theta}^2}{2}\,.  
\label{eq:raywod}  
\end{equation}  
This equation has the simple solution   
\begin{equation}  
\theta = \frac{2}{\lambda + {\lambda}_0} \,.  
\label{eq:Atheta1}  
\end{equation}  
After the absorption of the pulse the generators of the   
horizon have zero expansion, so we want $\theta = 0$ for   
$\lambda > 0$, so let   
\begin{equation}  
\theta(\lambda) =   
\left\{\matrix{0  \,, &  \,\, \lambda > 0 \cr  
2/(\lambda + {\lambda}_0) \,,   
&  \,\, \lambda < 0}\right. \,,  
\label{eq:Atheta2}  
\end{equation}  
i. e. , $\theta$ jumps discontinuously from $2/\lambda_0$ to   
$0$ as $\lambda$ passes through $0$. This gives us a   
$\delta$-function term in $d\theta/d\lambda$ of   
$-2 \delta(\lambda)/{\lambda}_0$. Thus the above form of   
$\theta$ is a solution of   
\begin{equation}  
\frac{d \theta}{d \lambda} = -\frac{ {\theta}^2}{2}   
- a \delta(\lambda) \,,  
\label{eq:Aray2}  
\end{equation}  
if $a = 2/{\lambda}_0$, or ${\lambda}_0 = 2/a$.   
(The divergence of $\theta$ at $\lambda = -2/a$   
represents a conjugate point in the distant past,   
which will be discussed later.)   
  
Let us now generalize our solution slightly by assuming that   
the pulse intersects the horizon not at $\lambda = 0$, but   
at $\lambda = \lambda_{+}$. Then our solution for $\theta$   
becomes   
\begin{equation}  
\theta(\lambda) =   
\left\{\matrix{0  \,, &  \,\, \lambda > \lambda_{+} \cr  
2/(\lambda - \lambda_{+} + 2/a) \,,   
&  \,\, \lambda < \lambda_{+}}\right. \,.  
\label{eq:Atheta3}  
\end{equation}  
 For the $\delta$-function   
pulse case, in the region where $\lambda < \lambda_{+}$,   
we can then integrate Eq.~(\ref{eq:thetaA})  
to get   
\begin{equation}  
{\rm ln} \, A = 2 \, {\rm ln}\,(\lambda-\lambda_{+} + 2/a) + C \,.  
\label{eq:AlnA}  
\end{equation}   
We fix the constant of integration $C$ by the requirement that   
$A \rightarrow A_f=$ final horizon area as $\lambda \rightarrow   
\lambda_{+}$, so we have   
\begin{equation}  
C = {\rm ln} \, A_f - 2 \, {\rm ln}\,(2/a) \,.  
\label{eq:C}  
\end{equation}  
Our expression for $A$ then becomes   
\begin{equation}  
A(\lambda) = A_f \, {\left(\frac{\lambda-\lambda_{+} + 2/a}{2/a}   
\right) }^2 \,, \,\, \lambda < \lambda_{+} \,,  
\label{eq:Alambda}  
\end{equation}  
or   
\begin{equation}  
A(v) = A_f \, {\left[\frac{a(e^{\kappa v} - e^{\kappa v_{+}}   
+ 2 \kappa)}{2 \kappa}   
\right] }^2 \,, \,\, v < v_{+} \,,  
\label{eq:Av1}  
\end{equation}  
where we have used Eq.~(\ref{eq:lambdaV}). We can write this   
in the form   
\begin{equation}  
A(v) =   
\left\{\matrix{A_f  \,, &  \,\, v \geq v_{+} \cr  
 A_f \, {\left\{\left[a(e^{\kappa (v-v_{+})} - 1) +   
2 \kappa \, e^{-\kappa v_{+}} \right] /  
(2 \kappa \, e^{-\kappa v_{+}}) \right\} }^2 \,,   
&  \,\, v \leq v_{+}}\right. \,.  
\label{eq:Av2}  
\end{equation}  
  
If $|v-v_+| \gprox {\kappa}^{-1}$, and since $v-v_+ \leq 0$,   
then $e^{\kappa (v-v_+)} \approx 0$, and we have   
\begin{eqnarray}  
A &\approx& A_f \, {\biggl[ { {2 \kappa\, e^{-\kappa v_+} -a} \over  
{2 \kappa\, e^{-\kappa v_+}} } \biggr]}^2 \nonumber \\  
&=& A_f \, {\biggl(1- \frac{a}{2 \kappa} \, e^{\kappa v_+} \biggr)}^2  
\equiv A_0 \,.  
\label{eq:Aapprox}  
\end{eqnarray}  
Thus on a timescale of order ${\kappa}^{-1} = 4M$, $A$ increases   
from $A_0$ to $A_f$. In the distant past (corresponding to   
$\lambda \rightarrow \lambda_+ -2/a$), $A \rightarrow 0$, as   
can be seen from Eq.~(\ref{eq:Alambda}), but this   
is the conjugate point associated with the formation of the black   
hole or with the past singularity in the case of an eternal   
black hole. After the black hole forms, its area remains nearly   
constant until a few times $M$ {\it before} the positive   
pulse arrives at $v=v_+$.   
  
We see that the horizon ``anticipates''   
the subsequent arrival of the pulse. This counterintuitive acausal   
behavior arises from {\it the way the horizon is defined}. Recall that   
the horizon is defined as the boundary of past null infinity, i. e.,   
the boundary of the region from which it is possible for light rays   
to escape to infinity (see, for example, Ref. \cite{WGR}, p. 300).   
Thus the horizon has the peculiar   
property that one must know the entire future evolution of   
the spacetime in order to know where the horizon intersects   
any given spacelike slice. For a static black hole, the null generators   
of the horizon have zero expansion. In order for the null generators   
in the horizon of the final (static) black hole to have zero   
expansion after the absorption of the pulse, the generators   
must have nonzero expansion prior to the pulse's arrival. The null   
rays in the horizon of the final black hole are rays that {\it would}   
have escaped to infinity were it not for the arrival of the   
positive pulse, which refocused them to have zero expansion.   
(See Fig. 59 of Ref. \cite{HE}, and the discussion   
in Ref. \cite{FN}.)  
  
Our calculation assumes that $\kappa$ is approximately constant   
and hence applies when the increase in area is small:  
\begin{eqnarray}  
A_f &=& A_0 \, {\biggl(1- \frac{a}{2 \kappa}  
 \, e^{\kappa v_+} \biggr)}^{-2}  \nonumber \\  
&\approx& A_0 \, {\biggl(1+ \frac{a}{\kappa}   
\, e^{\kappa v_+}  
 \biggr)}  \nonumber \\  
&=& A_0 + \Delta A \,.  
\label{eq:Aapprox2}  
\end{eqnarray}  
If we take $A_0=16 \pi {M_0}^2$ to be the initial area   
of the black hole in the distant past, where $M_0$ is   
its initial mass, then we have   
\begin{equation}  
\frac{\Delta A}{A_0} \approx \frac{a}{\kappa} \, e^{\kappa v_+}  
=2\,\, \frac{\Delta M}{M_0} \,.  
\label{eq:deltaAoverA0}  
\end{equation}  
In this approximation, the change in the mass of the black hole   
is   
\begin{eqnarray}  
\Delta M &=& \frac{a}{2 \kappa} \, M_0 \, e^{\kappa v_+}  
\nonumber \\  
&=& \frac{a \,A_0}{8 \pi} \, e^{\kappa v_+} \,,  
\label{eq:DM}  
\end{eqnarray}  
where we have used $\kappa = 1/(4M) \approx 1/(4M_0)$.   
This agrees with the result obtained by calculating the   
change in mass directly from Eq.~(\ref{eq:Mp}).  
  
Let us now check that the conjugate point which appears in our   
solution occurs in the distant past.   
  From Eq.~(\ref{eq:Atheta3}) we see that the   
conjugate point occurs at   
\begin{equation}  
\lambda = \lambda_c = \lambda_+ - \frac{2}{a} \,.  
\label{eq:conj}  
\end{equation}  
Using Eqs.~(\ref{eq:lambdaV}) and (\ref{eq:DM}),  
one can obtain   
\begin{equation}  
a= \frac{2 \, \Delta M}{M_0 \, \lambda_+} \,,  
\label{eq:a}  
\end{equation}  
and so   
\begin{equation}  
\lambda_c = \lambda_+ - \frac{M_0 \, \lambda_+}{\Delta M}   
= \lambda_+ \, \biggl(1- \frac{M_0}{\Delta M} \biggr) \,.  
\label{eq:lambdac}  
\end{equation}  
Since in our approximation $\Delta M \ll M_0$,   
$\lambda_c < 0$ and $|\lambda_c| \gg \lambda_+$.  
Note that $\lambda=0$,   
or equivalently $v=-\infty$ or $V=0$, corresponds to the   
intersection of the past and future horizons in the maximally extended
Schwarzschild spacetime. Thus for an eternal black hole, the conjugate
point cannot occur to the future of the past horizon. In the case of
a black hole formed from collapse, the same parameterization of    
$\lambda$ yields $\lambda > 0$ everywhere on the future horizon
outside of the collapsing body. In either case, $\lambda_c < 0$ 
implies that the conjugate point occurs in the distant past.

\end{document}